# Behavior of Ultrafine Particles in Electro-Hydrodynamic Flow Induced by Corona Discharge

Ravi Sankar Vaddi[1], Yifei Guan[2], Igor Novosselov[1,3]

[1]Department of Mechanical Engineering, University of Washington, Seattle, U.S.A.98195

[2]Department of Mechanical Engineering, Rice University, Houston, U.S.A.77005

[3] Institute for Nano-Engineered Systems, University of Washington, Seattle, U.S.A. 98195

## ABSTRACT

Ultrafine particle behavior in electro-hydrodynamic (EHD) flow induced by corona discharge is studied experimentally and numerically. The EHD flow serves as a primary particle aspiration/sampling mechanism, the collector does not require any additional flow generation. Multiphysics numerical model couples the ion transport equation and the Navier-Stokes equations (NSE) to solve for the spatiotemporal distribution of electric field, charge density, and flow field, the results are compared with experimental velocity profiles at the exit. The computed velocity and flow rate data are in good agreement with the experimental data; the maximum velocity is located at the axis and ranges from 1 m/s to 4 m/s as a function of corona voltage. Experimentally evaluated particle transmission trends for ambient and NaCl nanoparticles particles in the 20 nm - 150 nm range are in good agreement with the theoretical models. However, for particles in the 10 nm - 20 nm size range, the transmission is lower due to the increased particle charging resulted from their exposure to the high-intensity electric field and high charge density in the EHD driven flow. These conditions yield a high probability of particles below 20 nm to acquire and hold a unit charge. The transmission is lower for smaller particle (10 nm) due to their high charge to mass ratio, and it increases as the single-charged particles grow in mass up to 20 nm, resulting in their lower electrical mobility. For particles larger than 20 nm, the electrical mobility increases again as they can acquire multiple charges. The results shed insight into interaction of nanoparticle and ions in high electrical field environment, that occur in primary EHD driven flows and in the secondary flows generated by corona discharge.

## 1. INTRODUCTION

Gas-phase collisions between the particles and ion medium play an important role in governing the behavior of aerosols (Fuchs 1963, Marlow and Brock 1975, Zhuang, Jin Kim et al. 2000, Lee, Kim et al. 2016) and dusty plasmas (Pelletier 2000, Ravi and Girshick 2009). The presence of the electric field and the ion medium plays a major role in particle trapping since particles acquire a charge from ion collisions (Huang and Chen 2002). The electrostatic force on a charged particle in the electric field can be greater than gravitational, inertial and thermal forces. Electrostatic precipitator (ESP) devices can collect fine and ultrafine particles and are widely used in sampling and filtration applications. Conventional ESPs employ a point-plate, point-cylinder, wire-plate, point-ring configuration. In a typical electrostatic particle collector, the flow is induced by an external source the corona discharge induced flow is typically not considered. However, the ionic interaction used to collect the particles can be used to generate the flow and the enhance particle charging in corona discharge driven flow. It is challenging to gain insight into the particle charging mechanism in the high ion concentration and electric field environment due to the complexity of the physical phenomena and a lack of experimental data.

---

[1]ivn@uw.edu

Corona discharge is an electrical breakdown of air in which ions are generated in the high electric field region near the high energy anode, these ions drift towards the grounded cathode. The collisions of ions with the neutral air molecules result in a macroscopic wind, which is also known as electro-hydrodynamic (EHD) flow or ionic wind. The EHD effect has been used for plasma-assisted combustion (Starikovskii, Anikin et al. 2006, Ju and Sun 2015), convective cooling (Go, Garimella et al. 2006, Go, Garimella et al. 2007, Go, Maturana et al. 2008, Jewell-Larsen, Hsu et al. 2008) and control of the aircraft (Touchard 2008, Moreau, Benard et al. 2013). The application of EHD technology has been limited due to the modest pressure values achieved by the EHD blowers. However, in the applications with the low-pressure drop, the EHD driven flow can provide novel solutions (Jewell-Larsen, Parker et al. 2004, Drew, Contreras et al. 2017, Guan, Vaddi et al. 2018). Among the benefits of the EHD approach are the ability to operate at a small scale without moving parts and quiet operation (Moreau, Benard et al. 2013, Drew and Pister 2017, Dedic, Chukewad et al. 2019). The current-voltage relationship describes the ion transport between the electrodes. The classical voltage to the current relationship is derived by Townsend for a coaxial corona configuration (Townsend 1914). This quadratic relationship has been observed for other configurations, i.e., point to plate (Sigmond 1982) and point to ring corona (Guan, Vaddi et al. 2018). A generalized analytical model for voltage to current and voltage to velocity relationship for EHD driven flow has been recently developed (Guan, Vaddi et al. 2018). The maximum velocity for point-to-ring electrode configuration was recorded at ~9 m/s, the analytical model provides a good comparison with the experimental data.

Particle charging mechanisms have been an active research area, field and diffusion charging expressions (Rohmann 1923, Pauthenier and Moreau-Hanot 1932, White 1951, White 1963) were developed for large particles (0.3 μm - 10 μm). Fuchs (Fuchs 1947) and Marlow and Brock (Marlow and Brock 1975) developed diffusion charging expression for smaller particles, a combined field and diffusion charging expression was proposed by Liu and Kapadia (Liu and Kapadia 1978). Most particle charging expressions were developed for spherical particles, however recent experimental results for square particles demonstrated enhanced particle charging due to corners and edges (Unger, Boulaud et al. 2004). Experimental and numerical studies have demonstrated a good agreement in the size range of 0.3 μm - 10 μm (Goo and Lee 1997, Park and Kim 2000, Dau, Dinh et al. 2018). Multiple experimental studies for the PM in size range of 30 nm - 400 nm (Yoo, Lee et al. 1997, Miller, Frey et al. 2010, Dey and Venkataraman 2012, Roux, Sarda-Estève et al. 2016) agree with the theoretical models. Nanoparticle generation and behavior in low-temperature plasma produced by DBD discharge was studied (Jidenko and Borra 2005) and particles smaller than 20 nm are generated (Jidenko, Jimenez et al. 2007, Borra, Jidenko et al. 2015). Several researchers have also shown that for particles smaller than 30 nm, a fraction of particles was not charged and not collected (Pui, Fruin et al. 1988, Zhuang, Jin Kim et al. 2000, Li and Christofides 2006, Qi, Chen et al. 2008, Lin and Tsai 2010, Intra and Tippayawong 2011, Flagan and Seinfeld 2012). This phenomenon is called partial charging. Experimental and theoretical studies conducted by Dey et al. (Dey and Venkataraman 2012), Pui et al. (Pui,

Fruin et al. 1988), Li et al. (Li and Christofides 2006), Liu and Pui (Liu and Pui 1977) showed that Fuchs theory successfully predicted the charging probability of ultrafine particles. However, the scientific literature does not provide experimental data or numerical modeling related to the collection of nanoparticles in EHD dominated flow which is associated with high ion concentration and strong electric field.

In this manuscript, we analyze particle transmission in the primary needle-to-tube EHD flow. The flow is studied experimentally and by the numerical simulations to obtain the spatiotemporal characteristics of ion concentration, velocity, and electric field. Particles are aspirated by the corona discharge driven flow, charged due to their collision with ions and are collected on to the ground electrode. Nanoparticle transmission efficiency is determined experimentally at various corona voltages for ambient and NaCl particles showing low transmission (high collection) efficiency for particles below 20 nm. The experimental data suggest that the particles smaller than 20 nm can attain and hold a unit charge in the vicinity of the ionization region of the corona induced EHD flow, leading to their increased collection.

## 2. EXPERIMENTAL METHOD
### 2.1. Design and Working Principle of EHD Particle Collector

The EHD particle collector aspirates the particle into the corona induced flow, rapidly charges the particles in the charging region, and collects the charged particles on the ground electrode; no moving parts are required for operation as the flow is aspirated by the EHD phenomenon. The high ion concentration and the strong electric field between the corona and ground electrodes result in efficient charging and high collection efficiencies of particles. Fig. 1 shows the principle of operation of the EHD particle collector. The device consists of a high-voltage needle electrode positioned on the axis of symmetry and a grounded conductive tube serving as a collection electrode. When a high voltage is applied, the neutral air molecules are ionized by the strong electric field at the tip of the corona electrode (Townsend 1915, Sigmond 1982). In positive corona discharge, electrons are attracted to the high voltage corona electrode, positive ions such as $O_2^+$ and $O^+$ drift towards the cathode. As the high-velocity ions repelled from the corona electrode, they collide with the neutral air molecules driving the EHD flow. Particles aspirated by the EHD flow travel through the high electric field, high ion concentration (ion drift) region where high-velocity ions bombard the particle imparting a charge via two mechanisms: (1) diffusion charging which is due to random collisions and (2) field charging which is when the ions travel along the electric field. It is typically assumed that diffusion charging is predominant for smaller diameter particles, i.e., $d_p$ < 200 nm (Pui, Fruin et al. 1988, Hinds 1999). However, in ion-driven flow, the ion/molecule and ion/particle collisions are more frequent and more energetic than in the diffusion charging scenario. The Coulomb force caused by the electric field between the corona electrode and grounded collection substrate forces particles towards the collection electrode.

The EHD device used in this study consists of a corona needle and a ground collection electrode, as shown in Fig. 1. The high voltage needle is 0.5 mm thick tungsten wire with a tip curvature of 1 μm (measured using optical microscopy), the sharp tip yields high electric field strength and results in consistent EHD flow velocity data. As shown in the previous studies (Cheng, Yeh et al. 1981), needle sharpness affects the generation of the corona at lower voltages. The corona needle is regularly inspected for pitting using optical microscopy to ensure the consistent performance of the device. The ground electrode is an aluminum tube ID 7 mm with a rounded edge, the radius of curvature is 3 mm and tube length is 25 mm. The ground electrode has a rounded edge (radius of curvature 3 mm) to reduce the local E-field leading to spark-over, thus allowing the operation over wider voltage range. The electrode holder is fabricated using 3D printing from Polylactic Acid material (PLA). The needle is located on the axis of symmetry at 3 mm distance from the edge of the ground electrode, see Fig. 3.

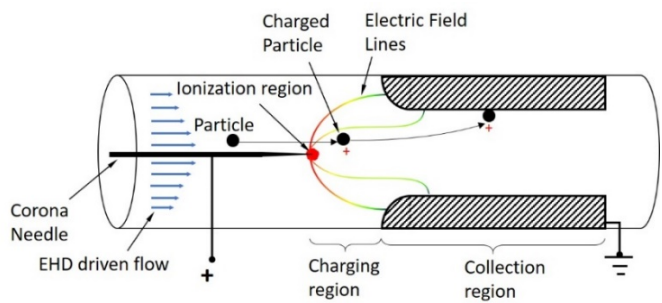

**Fig. 1.** Schematic of EHD particle collector in point to tube configuration.

### 2.2. Flow Field Measurements

The flow velocity is important in particle transmission study as it affects the particle residence time in the charging and the collection regions. Previous reports used an external pump to aspirate the particles through the corona region, the flow rate can be controlled to achieve maximum collection efficiency (Dixkens and Fissan 1999, Mahamuni, Ockerman et al. 2019). In the current work, particles are aspirated without the aid of an external pump and it is important to characterize the velocity generated by the corona discharge induced flow. A hot-wire anemometer (AN-1005) is used to measure the velocity profile at the outlet of the device. These velocity measurements are also used for calculations of the flow sampling rate. TSI 1213-20 hot wire probe connected to anemometer is positioned at the outlet of the device. The anemometer is calibrated for the range of 0.2 m/s - 5 m/s using the standard calibration procedure. The data from the anemometer is collected at a frequency of 10 kHz with a data acquisition module (National Instruments, myRIO-1900) for a sampling time of 10 seconds. A variable high voltage positive power supply (Bertan 205B-20R) is used to create the potential difference between the needle and the grounded tube. The corona current is measured on the cathode using a voltage drop across a 1 MΩ resistor as shown in Fig. 2(a). The onset of corona generation was observed at 2 kV; however, the current measurements in the experiments below 3 kV were not consistent in the day-to-day operations. In this work, the voltage on the needle is varied from 3 kV to 5 kV. For corona voltage above ~6 kV, spark over events occurred. All experiments were performed in ambient air at temperatures of 22 C - 25 C, relative humidity range of 30% - 35%, and pressure of 1 atm.

### 2.3. Test Particles and their Preparation

The transmission efficiency of the device is determined for two particle types (i) ambient particles from a typical laboratory environment, the particle chemical composition or their origin are not known and (ii) NaCl particles generated in the well-mixed aerosol chamber (He and Novosselov 2017, He, Beck et al. 2018). Both particles types have been previously used in electrostatic particle studies as test particles (Krichtafovitch I. A. 2005,

Miller, Frey et al. 2010, Roux, Sarda-Estève et al. 2016, Vaddi, Mahamuni et al. 2019). We estimated that the particles exceed their saturation charge due to the high charge density environment as they travel through charging region (as shown in Fig. 4) and a charge neutralizer is not used. A particle sizer (TSI SMPS 3910) is used to monitor the particle concentration. For ambient particles, their number concentration upstream was typically $3 \times 10^3$ count/cc with a median diameter of 45 nm. Previous reports have indicated that non-equilibrium low-temperature plasma generates small particles with diameter less than 20 nm(Jidenko, Jimenez et al. 2007, Borra, Jidenko et al. 2015). . We have performed a series of experiments to determine the particles that are generated from corona discharge. The particle concentration spikes 10 % and 5 % for 10 nm and 20 nm particles respectively. The concentration has dropped to the base line level after 10 seconds indicating that only a short burst of particles was produced in the onset of corona discharge as shown in SI Fig. 5. The background particles would not change the transmission efficiency results as the particle concentration reaches saturation.

In addition to the ambient particle experiments where the morphology and electrical properties of the particle may vary, the performance of the device is characterized using lab generated NaCl particles. The particles were generated with Up-Mist Medication nebulizer (MADA Products, Carlstadt, NJ, USA), using dilute solutions of NaCl in distilled water. The nebulizer is connected to HEPA-filtered air to provide the flow required for the generation of NaCl particles. The NaCl particles are generated in a custom 0.3 m$^3$ stainless steel, well-mixed aerosol chamber. The large volume of the chamber and the mixing fans provide well-mixed conditions, the aerosol concentration in the chamber was found to be spatially uniformed (He and Novosselov 2017). The particle size distribution depends on the solution concentration which was prepared to provide particles in size range of 10 nm–150 nm range. The sodium chloride particle concentration (~ 10$^6$ #/cc) is two orders of magnitude higher than the background concentration observed in the distilled water nebulization experiment (~1500 #/cc). During NaCl solution nebulization, the particle distribution is dominated by the NaCl particles, the size distributions for NaCl and distilled water are given in supplemental information, see SI Fig. 1 and Fig. 2

## 2.4. Experimental Setup for Transmission Efficiency

The experimental study characterizing the performance of the device consists of two parts: (i) determining the transmission efficiency for EHD flow as a function of corona voltage (ii) examining the effect of particle residence on the transmission efficiency where the flow rate was adjusted keeping the corona voltage constant. The measurements are carried out with the experimental setup as shown in Fig. 2(b) and (c). In the EHD flow scenario, see Fig. 2(b), two identical EHD devices are connected to the particle sizer in parallel. A selection valve allows for sampling from the EHD collector isokinetically or to switch to the parallel device and sample at the flowrates of the EHD collector. The flowrate on the EHD collector is determined from the velocity measurements. The reference particle concentration $C(no\_Voltage, d_p)$ was determined at the flowrate for each of the experimental conditions which were set by the sampling flow rate of particle sizer (0.8slpm) and an external pump connected in parallel and controlled by a needle valve. The flow rate of the reference flow is measured using a flowmeter (4140 D, TSI, Shoreview, MN). The flowmeter is calibrated aginst Gilibrator.

To investigate the effect of particle residence on the transmission efficiency, additional flow control was added to increase or reduce the particle residence time in the ionization and the collection regions while maintaining an active corona discharge. The device was connected by a T connector to the ultrafine particle sizer (TSI SMPS 3910) and an external pump with adjustable flow rate, see Fig. 2(c). Electrostatic dissipative tubing is used for fluidic connections to minimize particle losses. Both reference (no electric field) and EHD collector devices have identical geometries and fluidic connections, see Fig. 2(b). A particle sizer (TSI SMPS 3910) in the single size bin mode is used to measure the particle concentrations. The comparison of particle number concentration from the experiments with and without corona provides the transmission efficiency of the device. Similar methodology has been used in previous studies, e.g., (Huang and Chen 2002, Yao and Mainelis 2006) and it can be described by the expression.:

$$\eta = \frac{C(Voltage, d_p)}{C(no\_Voltage, d_p)} \;, \qquad (1)$$

where $\eta$ is the transmission efficiency, $d_p$ is the particle diameter indicating a specific particle size range, and $C$ is the particle concentration in the prescribed size bin. The transmission efficiency expression is similar to uncharged ratio as described in previous literature (Adachi, Kousaka et al. 1985, Romay and Pui 1992). In the current study, the data is recorded based on the measurement for individual size bin, rather the entire size spectrum scan to address temporal fluctuation in the particle concentration and size distribution in the environment. The sampling time was 60 seconds, and each experiment was repeated at least three times to obtain statistically relevant particle size data. The ozone concentration was measured using an ozone analyzer (Model 450, Teledyne Instruments) downstream (25 mm) of the tube over the range of corona voltages. Ozone concentration varies from 14 ppb – 24 ppb at the exit of the devise for an applied voltage 3 kV – 5 kV

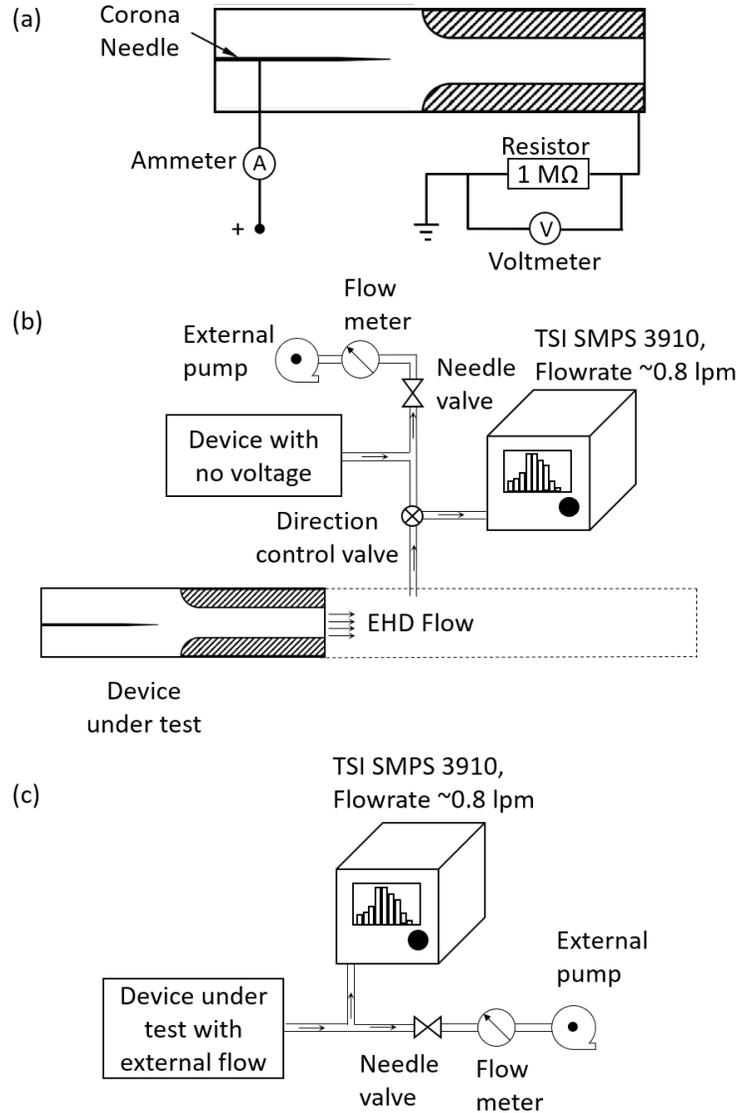

**Fig. 2.** Experimental setup for (a) corona current measurement and particle transmission study (b) EHD driven flow; the flow rate through the particle collector can be controlled with an external pump and (c) non-EHD experiment with active corona discharge

## 3. MODELING

Computational fluid dynamics (CFD) modeling is performed to gain insight into the flow properties in the EHD device as the velocity, ion concentration, and electric field are essential for studying the condition affect the particle behavior. Note that particle trajectories are not modeled in this work due to ambiguity in the particle charging model. Additional work is required to validate the particle charging and transport models. ANSYS Fluent software was used with custom subroutines for two-way coupling of ion motion and fluid flow. Fig. 3 shows the schematic of the modeled geometry. The model is taking advantage of axial symmetry and 2D axisymmetric model is used. The 3D simulation requires high computational cost considering high-resolution mesh requirements for the volumetric flux ionization model. The 2D assumption for modeling the corona region showed sufficient accuracy in previous work (Adamiak 2013, Guan, Vaddi et al. 2018) and the EHD flow general (Druzgalski, Andersen et al. 2013, Guan and Novosselov 2018, Guan and Novosselov 2019).

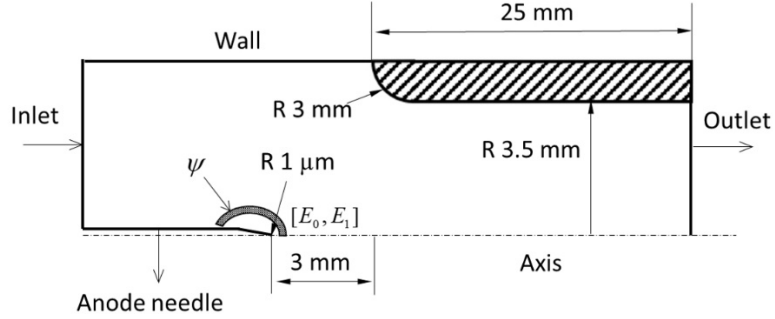

**Fig. 3.** Schematic of the computational domain; the model includes the ion generation region defined by the thresholds of the electric field

The flow field is solved using a finite volume laminar solver. The ion motion effects are incorporated by adding user-defined scalars to represent the electric potential $\varphi$ and charge density $\rho_e$. The electric force's effect on the flow is solved by introducing a body force $F_e = -\rho_e \nabla \varphi$ into the momentum equations, thus the governing equations for the flow are:

$$\nabla \cdot \mathbf{u} = 0 \qquad (2)$$

$$\rho \frac{D\mathbf{u}}{Dt} = -\nabla P + \mu \nabla^2 \mathbf{u} - \rho_e \nabla \varphi \qquad (3)$$

$\mu$ is the dynamic viscosity of the air, $\rho$ is the density of the air, $\mathbf{u}$ is the velocity vector and $P$ is the static pressure. The equations for charge transport are:

$$\frac{\partial \rho_e}{\partial t} + \nabla \cdot \left[ \left( \mathbf{u} + \mu_b \vec{E} \right) \rho_e - D_e \nabla \rho_e \right] = S_e \qquad (4)$$

$$\nabla^2 \varphi = -\frac{\rho_e}{\varepsilon_0} \qquad (5)$$

where $\mu_b$ is the ion mobility, which is approximated as a constant [2.0E-4 m²/(Vs)] and $\varepsilon_o$ is the electric permittivity of free space. $D_e$ is the ion diffusivity described by the electrical mobility equation (Einstein's relation):

$$D_e = \frac{\mu_b k_B T}{q} \qquad (6)$$

where $k_B$ is Boltzmann's constant ($\sim 1.381 \times 10^{-23}$ J/K), $T$ is the absolute temperature, and $q$ is the electrical charge of an ion, which is equal to the elementary charge ($1.602 \times 10^{-19}$ C). $S_e$ is the source term of charge density which has a unit of $C/m^3 \cdot s$, it is calculated from the corona current measured at the anode. In the simulation, the charges are introduced into the computational domain within the ionization zone boundary region at the rate calculated from the anode current. Instead of defining a thin surface within the computational domain to mark as the ionization zone boundary, a region with finite volume is determined by the electric field strength magnitude and constrained within 1mm of the needle tip.

$$S_e = \begin{cases} I/\psi, \text{ for } |E| \in [E_0, E_1] \text{ \& } x_{tip} - x < 1mm \\ 0, \text{ otherwise} \end{cases} \quad (7)$$

where $\psi$ is the volume of the region satisfying $|E| \in [E_0, E_1] \text{ \& } x_{tip} - x < 1\ mm$ and $I$ is the corona current. The $x_{tip} - x$ term limits the ion production along the needle; note that in the experiments, the needle tip extends only 1 mm from the needle holder. Additionally, the experimental observation shows that ionization zone is localized at the tip of the needle. The value of $E_0$ (*2.8 MV/m*). is the critical field below which the number of ions recombination is larger than production per drift length. The threshold $E_1$ is the breakdown electric field strength for air (*3.23 MV/m*). These ionization thresholds are used to mark the numerical "ionization region" where the charges (ions) are generated. More details on the method can be found in (Guan, Vaddi et al. 2018). Numerical schemes and boundary conditions are given in the supplemental information

## 4. RESULTS AND DISCUSSION

### 4.1. Voltage-Current Characteristics

The corona current and ion concentration at the exit are measured to determine the ion production and ion transport. Table I shows the corona current (anode current) vs. anode voltage. The current increases with the applied voltage quadratically, which agrees with other results in the literature for different corona configurations (Townsend 1914, Townsend 1915, Sigmond 1982, Giubbilini 1988, Kimio 2004). The current values from the experiments were used in the numerical model as the ionization zone boundary condition. In the CFD, the cathode (grounded electrode) current is determined by integrating the charge flux on the cathode surface.

The cathode current in the simulation agrees within 5% with the experimental measurements. Based on the simulations, the cathode recovers 85-90% of the ion current generated by corona, the other 10-15% are associated with ions exiting the geometry (See SI Fig. 3). These computed values of cathode current are in good agreement with the experimental data providing confidence in the numerical approach with respect to ion concentration field in the ionization and collection regions of the EHD collector.

$$I_{cathode} = \int_{\substack{cathode \\ area}} -\mu_b \rho_e \nabla \varphi \mathbf{dA}_{cathode} \tag{8}$$

where $I_{cathode}$ is the cathode current and $\mathbf{A}_{cathode}$ is the area vector of the cathode.

**Table I.** Comparison of cathode current between the experiments and CFD

| Voltage (kV) | Anode current (μA) | The experimental cathode current (μA) | CFD cathode current (μA) |
|---|---|---|---|
| 3 | 0.7 | 0.62 | 0.59 |
| 4 | 3.8 | 3.34 | 3.23 |
| 5 | 7.5 | 6.68 | 6.64 |

### 4.2. Flow Field Numerical Results

The numerical approach models the process by which the ion-molecule collisions accelerate the bulk flow. Fig. 4(a) shows the computed electric field lines. The maximum electric field strength is near the tip of the corona needle where a small radius of curvature concentrates the electric field lines and the field intensity reaches the threshold for ion generation. The effect of the space charge on the electric field is apparent by field line distortions in the region of high ion concentration. These distortions are significantly smaller away from the electrode tip where the charge density is reduced.

Fig. 4 (b) shows the ion density contours. The ions are generated at the needle tip as shown in SI Fig. 3, and their motion is dominated by the electric field due to their high electrical mobility, as the ion drift velocity is two orders of magnitude greater than the bulk flow (Sigmond 1982, Yu Zhang 2015, Guan, Vaddi et al. 2018). Downstream of the charging region, the electric field is weak, especially near the centerline, the ions exit the domain due to high flow velocities (see SI Fig. 3). A recirculation zone is formed upstream of the cathode tube near the rounded edge as shown in Fig. 4(c). This is due to the flow expansion which creates an adverse pressure gradient in the near-wall region.

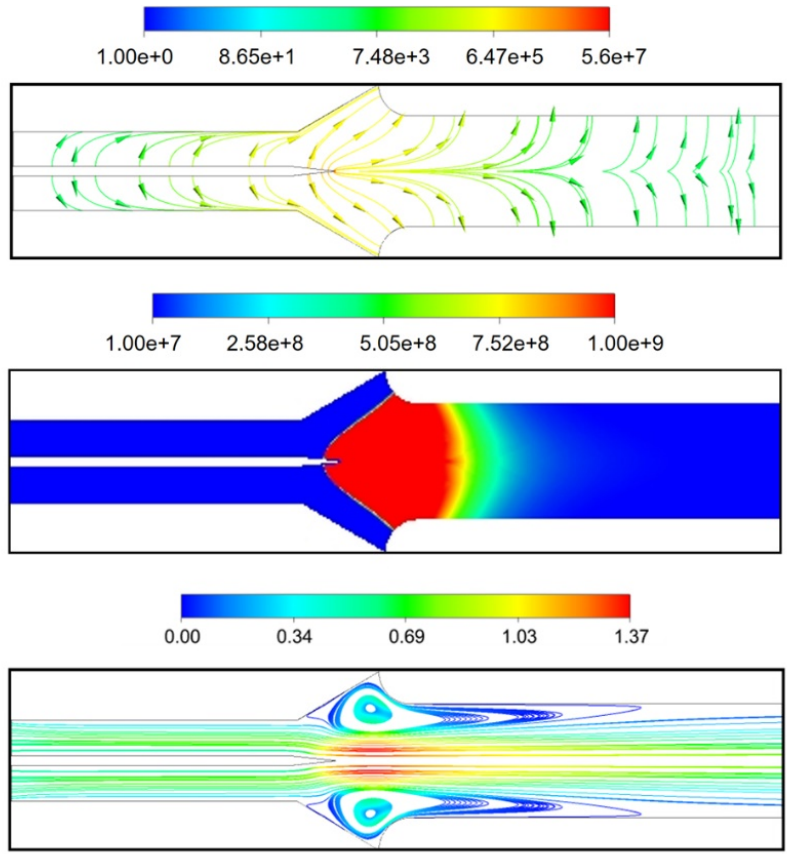

**Fig. 4.** Contour plots of the (a) electric field (V/m), (b) ion concentration (#/cc), the contours are clipped to 1e+9#/cc, maximum value is 5.93+9 #/cc (c) velocity (m/s) and for 3 kV corona voltage between the needle and the ground tube.

### 4.3. Velocity Voltage Characteristics

To validate the EHD modeling approach, the numerical results for corona voltages of $\varphi = 3$ kV - 5 kV are compared with the experimental exit velocities. Fig. 5 shows the velocity profiles plotted for three voltage values. The experiments and numerical results show the maximum velocity is located at the centerline; the profile decays with radial distance. The maximum velocity of the point-to-cylinder corona discharge device is ~4 m/s for both experiments and simulations at 5 kV corona voltage. At higher voltages arc discharge occurs, the flow velocity drops to zero. The maximum velocities in the numerical simulation are within 10% of the experimental data; the predictions are less accurate at the edges of the domain. The maximum outlet velocity increases linearly with corona voltage. The linear trend of centerline velocity is observed previously in experiments (Yu Zhang 2015, Guan, Vaddi et al. 2018). As the corona voltage increases, the discrepancy between the experiments and CFD increases and this may be due to averaging of velocity measurements across the hotwire element.

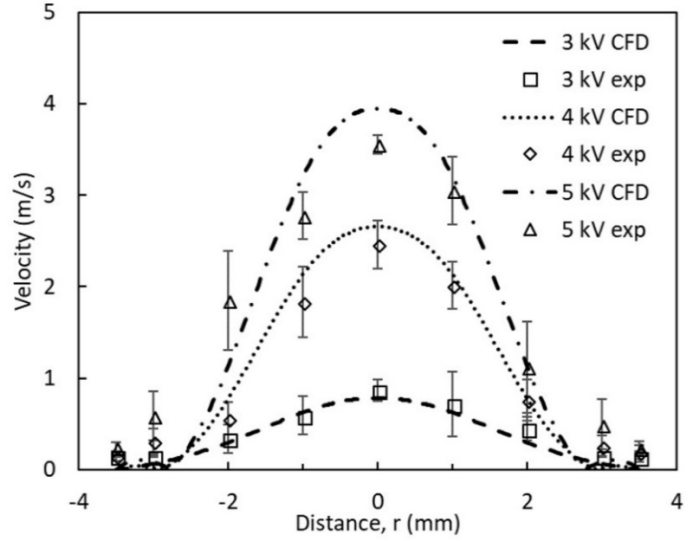

**Fig. 5.** Comparison of velocity profile between the experimental results and simulations at the outlet of the EHD induced flow device, as shown in Fig. 4(c).

The velocity profile shows that EHD induced flow in a point-to-tube corona discharge resemble Poiseuille flow near the axis and is significantly different from the pressure-driven flow profile near the walls. The point EHD source generates the flow similar to the submerged laminar jet flow (Landau 1959). Laminar flow characteristics are apparent from the experimental data. The Reynolds number ($Re$) is determined based on the tube diameter and the mean velocity at the exit; $Re$~160 for corona voltage of 3 kV and $Re$~400 for corona voltage of 5 kV. Since the 6 kV cases result in the arc, it appears that the corona induced flow without additional contribution from pressure term remains laminar for the considered internal flow geometry. If flow instabilities are present in the jet at its source, these temporal fluctuations decay by the time the flow reaches the outlet.

### 4.4. Particle Transmission

Particle behavior in the EHD flow was studied experimentally. Fig. 6shows the particle transmission efficiency of sodium chloride and ambient particles at different corona voltages. The lab generated NaCl particles have higher particle concentration compared to ambient particles. The transmission efficiency data is similar for both particle types. The transmission efficiency plot can be divided into three distinct regions (i) 10 nm – 20 nm, (i) 20 nm – 85 nm, and (1) 85 nm – 150 nm.

For all corona voltages, the transmission efficiency of particles smaller than 20 nm increases with the increase of particle size, i.e., the lowest transmission efficiency for the 10 nm – 20 nm range is observed for 10 nm particles. This behavior has not been previously investigated in the literature. The transmission efficiency of 10 nm particles decreases from 55% at 3 kV corona voltage to 30% for 5 kV corona voltage. These low transmission efficiencies indicate that in EHD flow 10 nm particle acquire charge with a higher probability that has been reported. Previous research (Fuchs 1947, Fuchs 1963, Zhuang, Jin Kim et al. 2000, Alonso, Hernandez-Sierra et al. 2003, Li and Christofides 2006, Lin and Tsai 2010) suggests that only a small fraction of particles is charged when the particle diameter is less than 30 nm. For example, according to classical diffusion charging models (Fuchs 1947, Pui, Fruin et al. 1988, Li and Christofides 2006), 12% - 37% of 10 nm particle would acquire charges by the thermal ions, and the contribution of the field charging is negligible for this particle size. In our experiments 45% - 70% of 10 nm particles were collected thus acquired at least one charge when passed through the charging region of the EHD driven flow. As the corona voltage increases, the ion concentration and the ion mobility (ion velocity) increases leading to more frequent and more energetic collisions with the particles. With respect to the increasing transmission efficiency in the 10 nm – 20 nm size range, the previous studies show that it is unlikely for these smaller particles to receive and hold multiple charges (Marlow and Brock 1975, Pui, Fruin et al. 1988, Lin and Tsai 2010) independent of particle type. As the particle size increases from 10 nm – 20 nm, their electrical mobility decreases resulting in the higher transmission efficiency. Similar trends have been observed in our experiments from 10 nm – 20 nm between NaCl and ambient particles for the range of applied voltages.

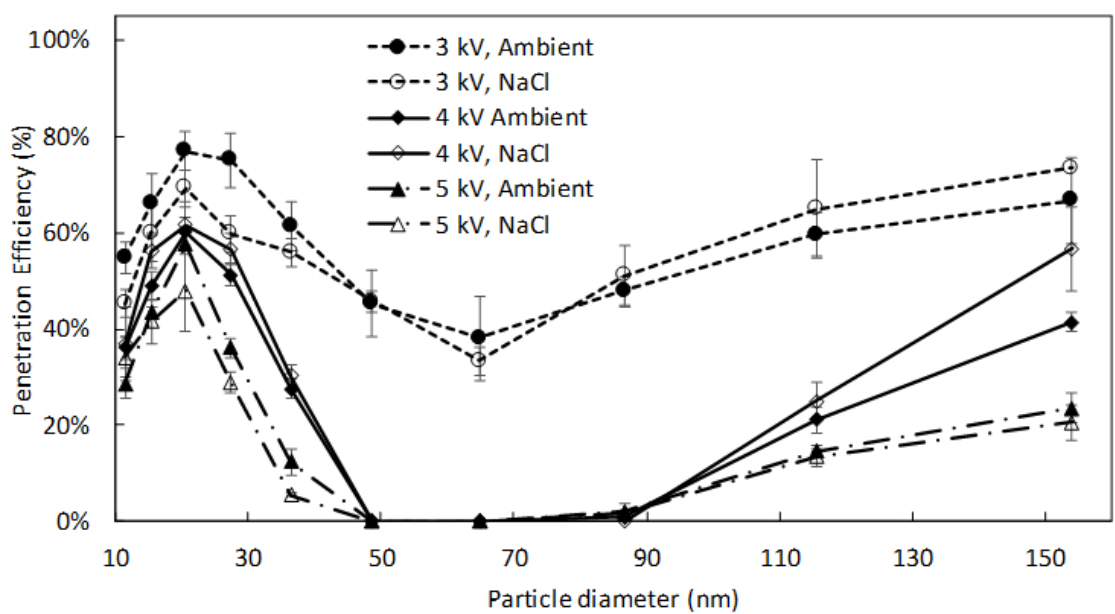

**Fig. 6.** Particle collection efficiency as a function of their size; results for NaCl (unfilled symbols) and ambient particles (filled symbols) at different corona voltages.

The particle size range of 20 nm – 85 nm exhibits a more traditional behavior; as the particle size increases the transmission efficiency decreases, due to the ability of the particles to carry multiple charges (Adachi, Kousaka et al. 1985, Pui, Fruin et al. 1988, Romay and Pui 1992, Zhuang, Jin Kim et al. 2000, Li and Christofides 2006) resulting in the higher electrical mobility thus, the lower transmission efficiency. Here the electrical mobility increases faster than the inertial and the drag forces governing the particle motion. For particle greater than 85 nm, the transmission efficiency increases with the increase of their diameter. The drag and inertial forces on the particle increase resulting in the decrease in the migration velocity even though particles attain multiple charges. This trend is consistent with the previous research showing that for polydisperse particles the transmission efficiency reaches a minimum and then increases for larger particles (Zhuang, Jin Kim et al. 2000, Li and Christofides 2006, Lin and Tsai 2010, Dey and Venkataraman 2012). However, one of the key findings presented in the current work is different the ratio of charged to uncharged particle in 10-20 nm region. We demonstrate that for 10-20 nm particle the ratio is approaching unity, as all particles are collected in the presence on repelling electrode, thus all particles must possess one or more charges. At this time, we do not have a way to quantify the number of charges as a function of particle size, particle morphology or its chemical composition.

The particle-laden flow passes through the charging region where both ion concentration and electrical field are high resulting in the high collision frequency between the ions and the particle in the flow. The collisions with high energy ions result in high particles charging efficiency and lower particle transmission. The highest charging rate is at the tip of the electrode as the ion concentration (2.44E+11 #/cc), and electric field strength (7.49E+07 V/m) are the highest. The maximum ion concentration and electric field strength for different corona voltages are given in Table II. The ion concentration reduces away from the tip due to radial ion motion caused by the ion drift towards the ground electrode and the space charge effect.

**Table II**. Maximum computed ion concentration and electric field strength in the ionization zone a function of corona voltages.

| Voltage (kV) | Ion concentration (#/cc) | Electric field strength (V/m) |
|---|---|---|
| 3 | 5.93 E+09 | 5.6E+07 |
| 4 | 8.62E+10 | 6.78E+07 |
| 5 | 2.44E+11 | 7.49E+07 |

To summarize, the trends of particle transmission in EHD driven flow is similar to the previously reported results for particle greater than 20 nm. However, the significantly lower transmission of 10 nm particles is observed likely due to the efficient charging in the region of high ion concentration / high electrical field within the corona discharge. Here, we also do not to quantify the exact number of charges on the particle as a function of ion concentration or electrical field strength; the detailed analytical or empirical model for the dynamic particle charging process in the corona region is not available. The charges acquired on the particle surface may be stripped from the particle by collision with neutral molecules as the particle travels through the domain into the region with less intense E-field with lower charge density.

To gain insight into the particle charging and capture dynamics, a series of experiments and numerical simulations were performed (i) by varying the particle residence time in the charging zone and (ii) by varying the particle mobility in the collection zone. The flow rates (thus the residence times) are controlled by the external pump as shown in Fig. 2(a). Though the flow in these experiments is not driven by EHD, all particles travel through the ion drift region. The residence time is a function of the bulk flow rate as well as the local flow field effect that is affected by the addition to the body force generated by the corona discharge. Fig. 7 shows the results of the numerical simulations. The baseline case is the EHD driven flow at 5 kV and flow rate of ~2 slpm (U = 0.9 m/s, $Re$~400) as determined by both CFD and by integrating the experimental velocity profile. Two additional cases are examined where the flow rates were set to 1 slpm (U=0.45 m/s, $Re$~200) and 5 slpm (U=2.25 m/s, $Re$~ 1000) to investigate the effect of the residence time. The corona induced flow has a significant effect on the velocity profile. Fig. 7 (right) shows flow streamlines colored by the non-dimensional parameter $X$ defined as the ratio of electrostatic force to the inertial force $X = \rho_e \varphi / \rho \mathbf{u}^2$ (Guan, Vaddi et al. 2018). As the flow rate increases, the inertial term contribution acting on the flow and the particles increases as shown by the smaller region of $X > 1$. For the 5 slpm case, this EHD dominated region exists only near the needle tip while for the lower flowrates all streamlines (thus the particle entering the device) experience $X > 1$ condition.

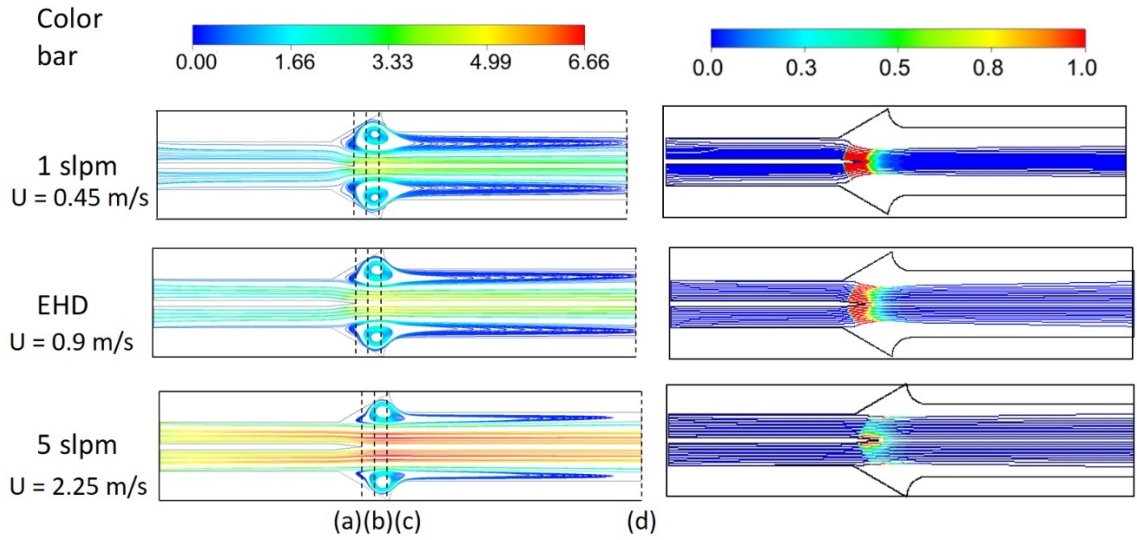

**Fig. 7.** Left: Velocity streamlines colored by velocity magnitude. Right: non-dimensional parameter $X$ (right) plotted on path lines for 1 slpm, EHD (~2 slpm), and 5 slpm flowrates. The dash lines indicate the location at which velocity profiles are compared (See SI Fig. 4).

Fig. 8 shows the transmission efficiency of ambient particles for 1 slpm (U = 0.45 m/s), EHD (U = 0.9 m/s), and 5 slpm (2.25 m/s) cases, the corona voltage for all cases is 5 kV. The transmission efficiency trend for particle greater than 20 nm is similar to the data as shown for EHD cases (see Fig. 6). For the 10 nm - 20 nm size range, the trends change as a function of the flow rate. As expected, the low flow rate (high residence time) results in lower transmission for all particles. The trends similar to EHD flow is observed for particles in 10 nm - 20 nm range suggesting the high fraction of the particles are charged, and these particles have sufficient time in the electric field to be collected onto the ground electrode. The transmission efficiency for 10 nm decreased from 29% to 15%. Another important data trend is the decrease in transmission of 20 nm particle from ~60% in the EHD case to ~20% in the 1 slpm case. If it is assumed that 20 nm particle can carry only a single charge, the most likely explanation for this drop the transmission efficiency is the increase of particle residence time in the high E-field region. The estimation of the particle residence time distribution is challenging as the flow profile is non-uniform, strong recirculation patterns exist in the particle charging region due to the local momentum source. The higher flow rate case is dominated by the pressure-driven flow as indicated by the parameter $X$. As the residence time decreases, the transmission efficiency of 10 nm particle increases from 29% to 75% and decreases for the sizes up to 85 nm and then increases, which similar to previously reported results (Zhuang, Jin Kim et al. 2000, Alonso, Hernandez-Sierra et al. 2003, Li and Christofides 2006, Lin and Tsai 2010, Dey and Venkataraman 2012).

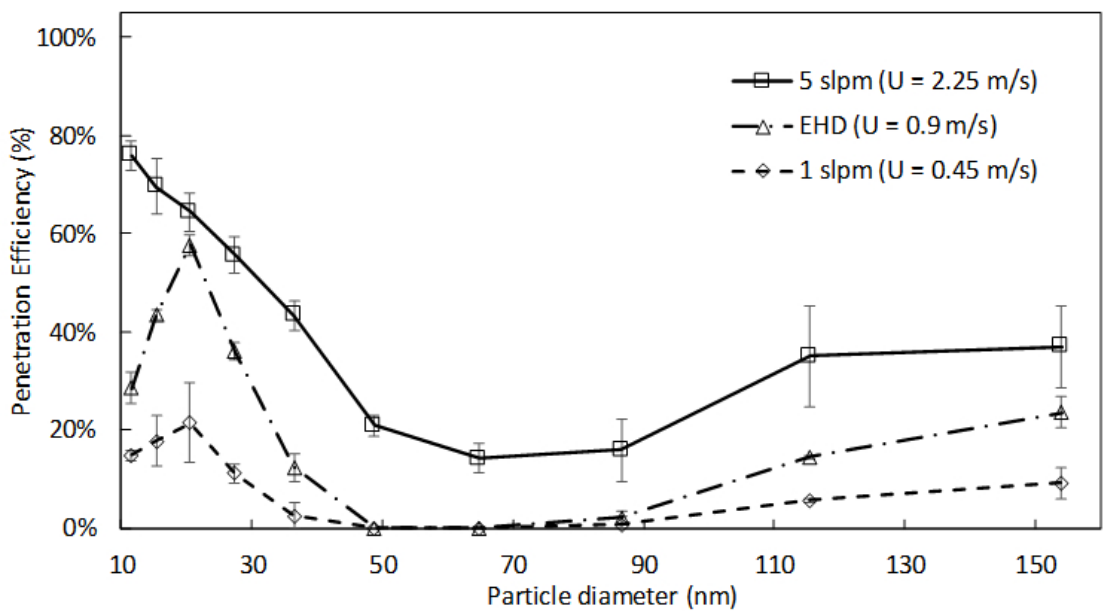

**Fig. 8.** Particle transmission efficiency as a function of particle size and flow rate for corona voltage of 5 kV

In the EHD driven case, the flow residence time in the region dominated by EHD ($X>1$) is ~8 ms -10 ms. As the flow rate increases to 5 slpm, the residence time drops to ~ 2 ms -3 ms, and it increases to ~ 17 ms – 20 ms for 1 slpm. The residence time of particle aspirated by the flow is not calculated due to challenges related to the particle charging models in the EHD driven flow. However, based on the experimental results, the fraction of 10 nm - 20 nm particles acquiring charge is as high as 80% for the case with the largest EHD dominated region. The EHD dominated region is also varying in size, thus, a greater portion of the flow (and the particles) passes through it at the lower velocity cases.

The particle forcing in the collection region is varied by introducing a repelling electrode along the axis of symmetry, the operation principle and design details are shown in SI Fig. 6. A voltage of 100 V is applied to the repelling electrode, to increase the Coulombic force acting on the particle in the region between the repelling electrode and the collection tube. The applied voltage on the repelling voltage have no effect on the ionization current (i.e., ion concentration in the charging did not change). Fig. 9 shows the penetration efficiency of particles with and without repelling voltage. For all the particles, the penetration efficiency decreases with the addition of repelling voltage. The transmission efficiency of 10 nm particles decreases from 28 % to 3 % and for 20 nm particles decreases from 58 % to 3.5

%. Similar results have been observed in previous reports (Huang and Chen 2002, Mahamuni, Ockerman et al. 2019, Vaddi, Mahamuni et al. 2019). These low transmission efficiencies indicate that all the particles receive and retained positive charge(s) in the high ion concentration region. As shown in Fig. 7, due to mass conservation in the flow acceleration zone, the majority of stream lines are forces to pass through the high ion concentration region. The ion concentration in the ionization region exceed values of 10E+9 #/cc, which are higher than reported in literature (SI Fig. 3 and Table II). Though additional studies to separate the effects charging and residence time in the high-intensity E-field are needed, one can conclude that the charging and collection of the ultrafine particles can be enhanced by their exposure to high charge density – high electric field region. The detailed charging mechanism is not considered in this work, however, the information presented here can aid the development of modified particle charging and transport models that can account for the effect of high ion concentration and strong E-field on particle transport.

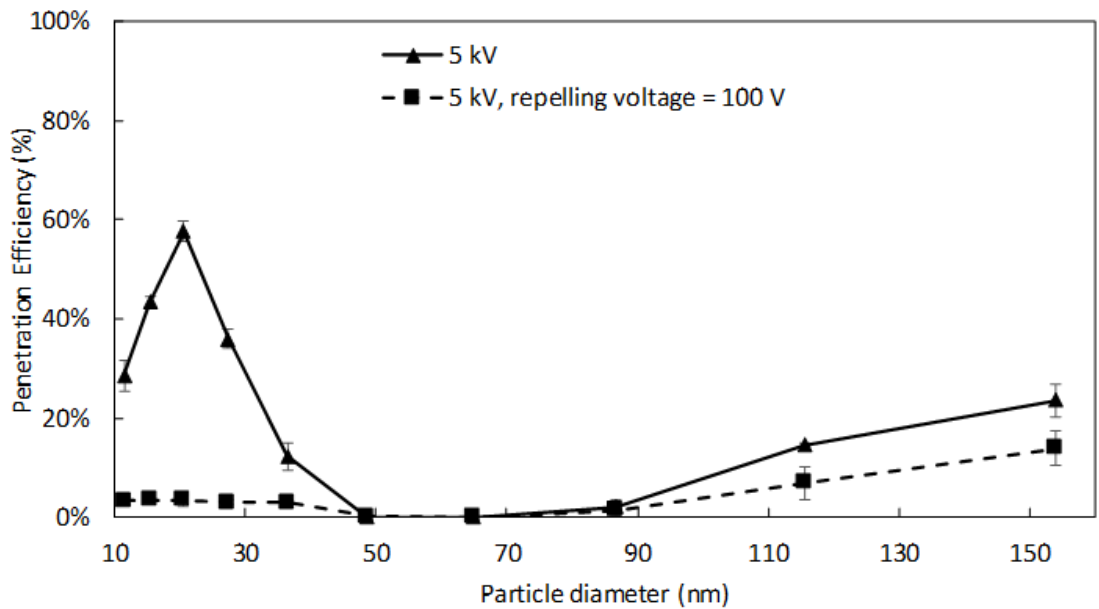

**Fig. 9.** Particle transmission efficiency as a function of particle size and repelling voltage for corona voltage of 5 kV

## 5. CONCLUSION AND DISCUSSIONS

This paper provides an experimental and numerical investigation of particle behavior in the EHD flow. An EHD needle-to-tube device aspirates the flow and collects the particle sample onto the collection electrode without the use of external pumps or any moving parts. The experimental data includes voltage, current, exit velocity profile measurements. Multiphysics numerical simulations show the interaction of the Coulombic force exerted by the ions on the airflow. The addition of charge flux as a model for the gas ionization zone allows for the direct computation of EHD flow adding the body force to the modified NSE.

The numerical simulations agree with experimental data within 10%. The corona induced flow for the investigated internal flow scenarios remains laminar, the *Re* = 100-400 for the range of operating corona voltages.

Ambient PM and NaCl nanoparticles were used to study the particle behavior in EHD-driven flow, the transmission efficiency is independent of particle type. Measured transmission efficiencies in EHD device are in good agreement with the traditional theories except for the particles in 10 nm - 20 nm range, the particle transmission for flow for EHD driven flow is significantly lower. The transmission efficiency for smaller particles is lower than reported by previous research likely due to the high fraction of 10 nm - 20 nm particles acquiring a unit charge in the EHD dominated region (*X>1*). As the particle size increases from 10 nm to 20 nm, their electrical mobility reduces due to the increase in particle mass while still possessing only a single charge. This hypothesis is further tested by varying the particle residence time and particle mobility in the EHD dominated region.. The transmission efficiency drops to 15% - 20% indicating that the fraction of 10 nm - 20 nm particles with at least one charge is greater than 80% - 85% when the particle residence time increases. The charging to uncharged particle ratio is approaching unity in the particle mobility experiments, as all the particles are collected in the presence of repelling electrode, thus all particles must possess one or more charges. These results suggest the charging of nanoparticles can be enhanced by their prolonged exposure to ion bombardment in the high charge density, high electric field region.

## 6. ACKNOWLEDGMENTS

This research was supported by the National Institutes of Health grant NIBIB U01 EB021923 and NIEHS R42ES026532 subcontract to the University of Washington.

## NOMENCLATURE

| | |
|---|---|
| $\eta$ | Collection efficiency of the particle collector |
| $d_p$ | Particle diameter (nm) |
| $C$ | Particle concentration (#/cc) |
| $\varphi$ | Electrical potential (V) |
| $\rho_e$ | Charge density (C/m$^3$) |
| $F_e$ | Electrostatic body force (Pa) |
| **u** | Velocity vector (m/s) |
| $\mu$ | Dynamic viscosity of air (kg/ m-s) |
| $\rho$ | Density of air (kg/m$^3$) |
| $P$ | Static pressure (Pa) |
| $\mu_b$ | Ion mobility (m$^2$/V-s) |
| $E$ | Electric field (V/m) |
| $\varepsilon_o$ | Electric permittivity of free space (F/m) |
| $D_e$ | Ion diffusivity (m$^2$/s) |
| $S_e$ | Source term for charge density (C/m$^3$-s) |
| $\psi$ | Ionization volume (m$^3$) |
| $I$ | Anode current (μA) |
| $[E_0, E_1]$ | Electric field criteria limits for ionization boundary (V/m) |
| $I_{cathode}$ | Cathode current (μA) |
| $A_{cathode}$ | Area vector of the cathode |
| $X$ | Non-dimensional parameter for the ratio of electrostatic force to inertial force |
| $Re$ | Reynold number |
| $R$ | Radial dimension (mm) |
| $U$ | Average velocity (m/s) |

# Supplemental Information

## 1. Computational Parameters

Table I shows the numerical schemes used in the CFD calculations. The second order upwind scheme is used to reduce numerical diffusion. The transient laminar solution is computed, the convergence criteria and the simulation time are set to achieve time steady velocity profile at the outlet. Since the ions drift velocity is orders of magnitude greater than the convective flow velocity, the solution for charge transport and electric field converge significantly faster (in convective time) than the flow equations. The boundary conditions are shown in Table II. The total pressure difference between the inlet and outlet is zero as the flow is accelerated only by the ionic drag.

**Table I.** Numerical schemes

| Model Parameter | Spatial Discretization |
|---|---|
| P-V Coupling | SIMPLE |
| Pressure | $2^{nd}$ order upwind |
| Momentum | $2^{nd}$ order upwind |
| Electric potential | $2^{nd}$ order upwind |
| Charge density | $1^{st}$ order upwind |

**Table II.** Boundary conditions for the numerical simulations

| Boundary | The value given at the boundary |
|---|---|
| Inlet pressure | Atmospheric pressure |
| Outlet pressure | Atmospheric pressure |
| Anode needle | 3~5 kV & Zero diffusion flux for charge |
| Cathode tube | 0 kV & Zero diffusion flux for charge |
| Wall boundaries | Zero diffusion flux for electric potential & charge density |

## 2. Particle Concentration Data

Two sets of experiments were performed in this study to understand the particle generation from (i) NaCl in distilled solution and (ii) distilled water. Aerosolization of the solutions using MADA Up-Mist™ Medication nebulizer (MADA Products, Carlstadt, NJ, USA) generated particles in the range of 10-150 nm. The experiments were performed in a custom 0.3 m³ stainless steel, well-mixed aerosol chamber. The large volume of the chamber with mixing fans provides well-mixed conditions, the aerosol concentration in the chamber was found to be spatially uniformed with the operation of the mixing fans (He and Novosselov 2017). The sampling time was 60 seconds, and each experiment was repeated three times to obtain statistically relevant particle size data. All the experiments have been performed at a constant relative humidity of 35%. Fig. 1 shows the particle number density in the distilled water nebulization experiments; the total concentration in the chamber is ~ 1500 #/cc. Fig. 2 shows the particle spectra for sodium chloride solution. The particle density for sodium chloride solution experiment is two orders of magnitude greater than during the distilled water nebulization. During NaCl solution nebulization the particle distribution is dominated by the NaCl particles.

---

[1]ivn@uw.edu

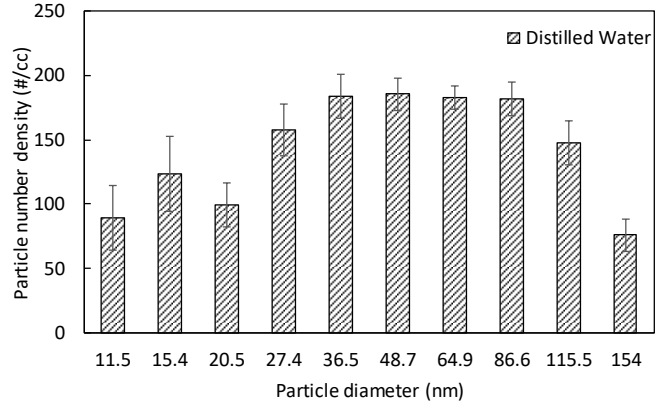

**Fig. 1.** Particle number density (#/cc) distribution for distilled water. Majority of the particles are in the size range of 36- 86 nm

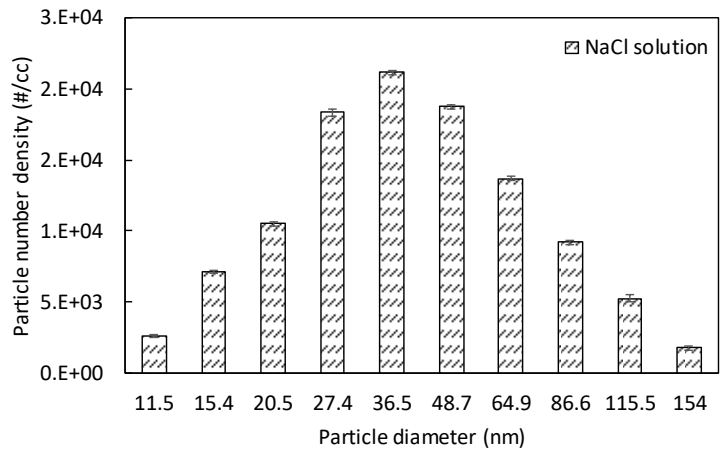

**Fig. 2.** Particle number density distribution (#/cc) in sodium chloride experiments.

### 3. Ion Concentration Data

The ions are generated at the needle tip as shown in Fig. 3 (a), and their motion is dominated by the electric field due to their high electrical mobility, as the ion drift velocity is two orders of magnitude greater than the bulk flow. Downstream of the charging region, the electric field is weak, especially near the centerline, the ions exit the domain due to high flow velocities. The cathode recovers 85-90% if the ion current generated by corona, the other 10-15% are associated with ions exiting the geometry as shown in Fig. 3 (b)

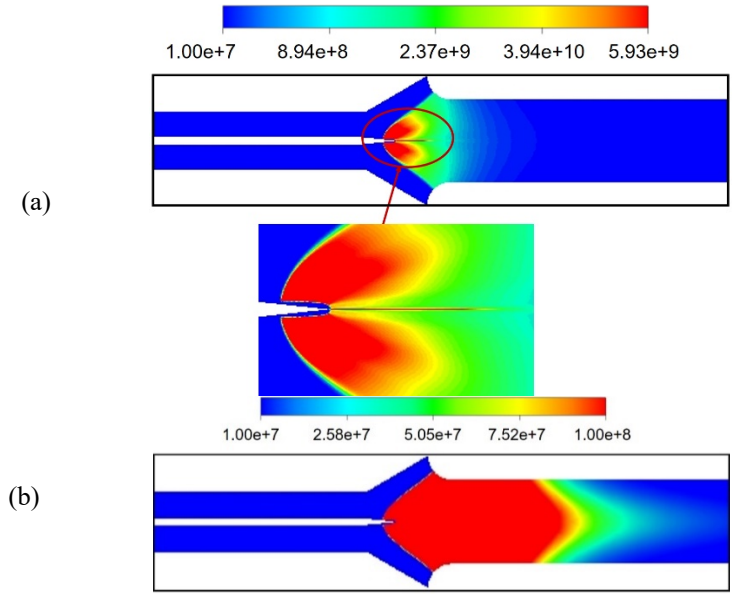

**Fig. 3.** Contour plots of the ion concentration (#/cc) (a) at the needle tip, maximum value is 5.93e+9 and (b) ions exiting the outlet. the contours are clipped to 1e+8 #/cc.

## 4. Velocity Profile Data

The velocity profiles at the cross-sections in the charging zone and at the exit are studied to elucidate the residence time effect on particle collection. Fig. 4(a) shows the velocity profiles at the anode for three different flow rates at a fixed voltage of 5 kV. The higher flow rate reduces the time for the particles to acquire charges as well as the charged particles residence time in the high-intensity electrical field (once the particle sufficiently enters grounded tube the electrical field drops significantly. Both of these conditions yield higher penetration efficiency. Fig. 4 (b) and (c) shows the velocity profile at the cathode and in between the electrode pair. The velocity profiles have a recirculation region near the wall, the length of the recirculation increases for low flow rates indicating the greater influence of the EHD effect. The corona driven flow produces an adverse pressure gradient at the wall due to local flow acceleration at the axis. For the low flow rate cases, particles entering the device may get trapped in the recirculation regions which increases their residence time in the charging region decreasing their penetration efficiency. However, in the laminar flow, cross-stream transport is slow (governed by molecular diffusion), and the fraction of total flow entering the recirculation cannot be very high.

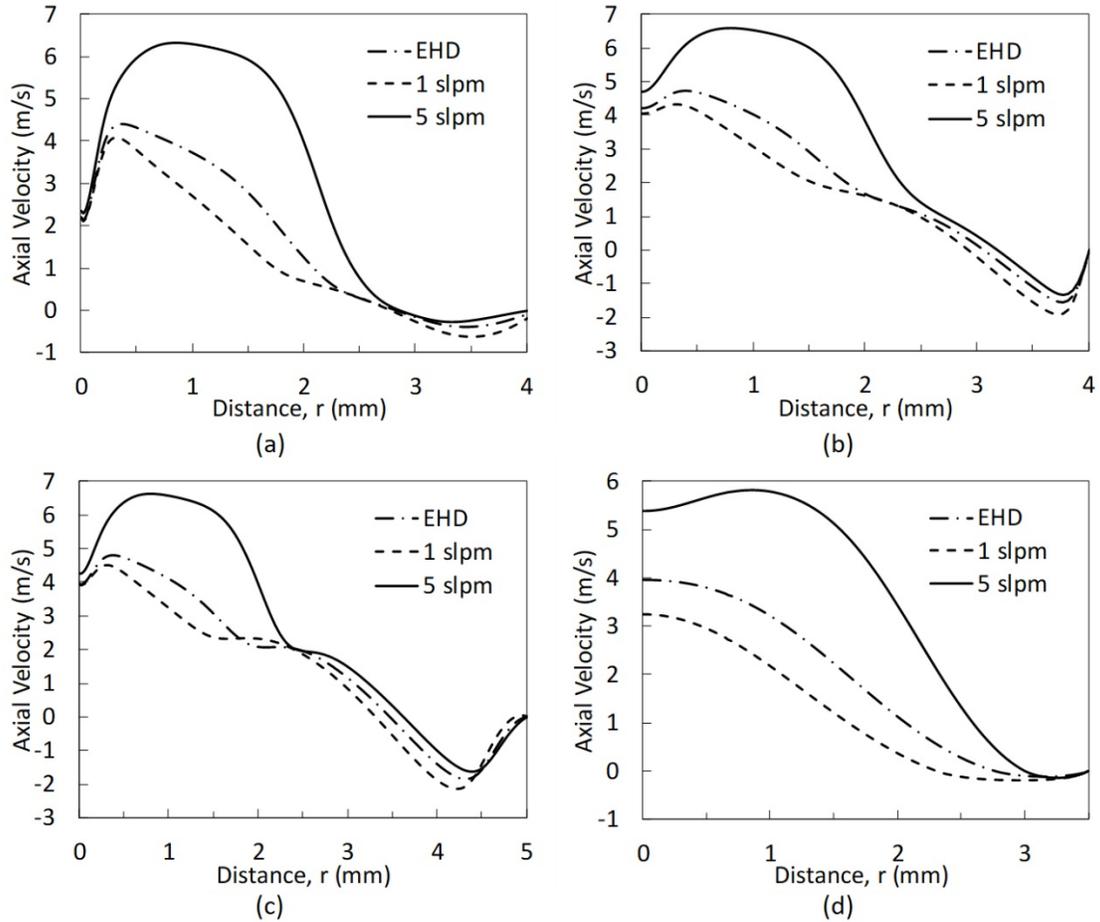

**Fig. 4.** Computed velocity profiles in the EHD collector at various axial cross-sections for 5 kV potential difference between the electrodes: a) at the anode electrode (tip of the needle), b) at the cathode electrode (plane aligned with edge of the tube), c) halfway distance between anode and cathode, d) at the exit

## 5. Background Particle Concentration

Previous reports have indicated that non-equilibrium low-temperature plasma generates small particles with diameter less than 20 nm (Borra, Jidenko et al. 2015). To evaluate particle possible particle generation in our experiments were performed a series of experiment. A particle sizer (TSI SMPS 3910) is used to monitor the particle concentration. The experiments were performed in a custom 0.3 $m^3$ stainless steel, well-mixed aerosol chamber. The sampling time was 60 seconds, and each experiment was repeated three times to obtain statistically relevant particle size data. The experimental study characterizing the particle background consists of two parts (i) determining the background particle concentration when the plasma is "OFF" and (ii) determining the background particle concentration for the plasma "ON" conditions. The time series of particle concentration are shown in Fig. 5. The background particle concentration is measured for 10 seconds and then the plasma is turned ON and the measurements are carried out. The particle concentration increases when the plasma is turned on but reaches to equilibrium in 10 seconds.

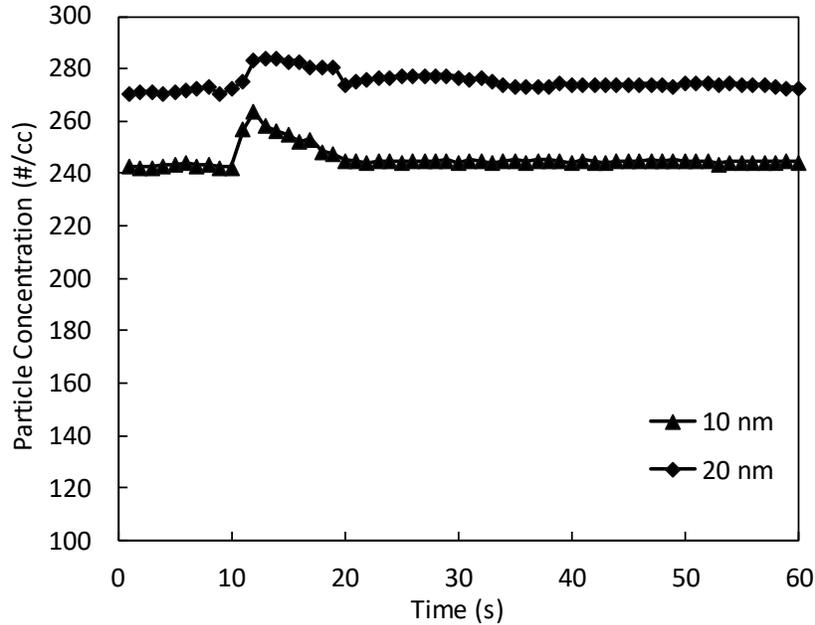

**Fig. 5.** Time fluctuations of particle background concentration for corona discharge

## 6. Design of Dual Stage EHD Particle Collector

The dual stage EHD collector is divided into two regions: the ionization region and the collection region. Fig. 6 shows the principle of operation of the dual stage EHD particle collector. The ionization region consists of a high-voltage needle electrode positioned on the axis of symmetry and a grounded conductive tube serving as a collection electrode. When a high voltage is applied, the neutral air molecules are ionized by the strong electric field at the tip of the corona electrode. In positive corona discharge, positive ions drift towards the cathode and the high - velocity ions collide with the neutral air molecules driving the EHD flow. Particles aspirated by the EHD flow travel through the high electric field, high ion concentration (ion drift) region where high-velocity ions bombard the particle imparting a charge. The Coulomb force caused by the electric field between the corona electrode and grounded collection substrate forces particles towards the collection electrode. The Coulomb force on the particles is increased by placing an additional electrode with positive polarity and this is called as repelling electrode.

The dual stage EHD device used in this study consists of a corona needle and a ground electrode for flow aspiration, as shown in Fig. 6. The high voltage needle is 0.5mm thick tungsten wire with a tip curvature of 1 μm (measured using optical microscopy), the sharp tip yields high electric field strength and results in consistent EHD flow velocity data. The ground electrode is an aluminum tube ID 7 mm with a rounded edge, the radius of curvature is 3mm; tube length is 25mm. The repelling electrode is 0.4 mm dia rod; length is 16 mm and it is placed up to 2/3$^{rd}$ lengths of the ground tube. The electrode holder is fabricated using 3D printing from Polylactic Acid material (PLA).

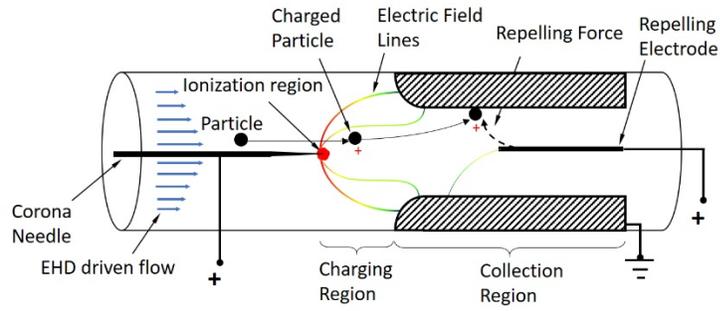

**Fig. 6.** Schematic of dual stage EHD particle collector in point to tube configuration